\documentclass[letterpaper,twocolumn,10pt]{article}
\catcode`\D=12

\catcode`\D=11
\PassOptionsToPackage{hyphens}{url}
\PassOptionsToPackage{breaklinks, pdfborder={0 0 0}}{hyperref}
\usepackage[dvipsnames]{xcolor}
\usepackage[style=base]{caption}
\usepackage[10pt]{sigmin}
\usepackage[a-1b]{pdfx}
\usepackage{hyperref}
\hypersetup{citecolor=blue,linkcolor=blue}
\usepackage[square,comma,numbers,sort&compress]{natbib}
\usepackage{subcaption}
\usepackage{times}
\usepackage[T1]{fontenc}
\usepackage{mathptmx}
\usepackage{relsize}

\usepackage{microtype}
\usepackage{pftools}
\usepackage{pifont}
\usepackage{fancyhdr}
\usepackage{xfp}
\usepackage[12hr]{datetime}
\usepackage{amsthm}
\input{glyphtounicode}
\usepackage{tabularx}
\newcolumntype{C}{>{\centering\arraybackslash}X}
\newcolumntype{L}{>{\raggedright\arraybackslash}X}
\newcolumntype{R}{>{\raggedleft\arraybackslash}X}
\newcolumntype{j}{>{\hsize=.65\hsize\arraybackslash}X}
\newcolumntype{h}{>{\hsize=.3\hsize\arraybackslash}X}
\newcolumntype{k}{>{\hsize=.65\hsize\raggedleft\arraybackslash}X}

\usepackage{tikz}
\usetikzlibrary{arrows,arrows.meta, shapes.geometric,decorations.pathreplacing,calc,positioning,matrix,tikzmark}
\usepackage{amsmath}
\usepackage{amssymb}
\usepackage{hyperref}
\usepackage{xspace}
\usepackage{booktabs}
\usepackage{multirow}
\usepackage{verbatim}
\usepackage{listings}
\usepackage{fancyvrb}
\usepackage[utf8]{inputenc}
\usepackage{enumitem}
\usepackage[group-separator={,}]{siunitx}
\usepackage[inline,final]{showlabels}
\newif\ifdraft
\draftfalse 

\newcommand{\insertnote}[3]{\ifdraft{\noindent\textcolor{#1}{\textbf{#2:} #3}}\fi}

\newcommand{\fk}[1]{\insertnote{red}{FK}{#1}}

\def\Snospace~{\S{}}

\input{glyphtounicode}
\bibpunct[: ]{[}{]}{,}{n}{XXX}{XXX}

\newcommand{\cc}[1]{\mbox{\texttt{\detokenize{#1}}}}

\newcommand{\sigmaos}{\textsc{$\sigma$OS}\xspace}

\newcommand{\xcontainers}{$\sigma$containers\xspace}

\newcommand{\Sandbox}{Initscript\xspace}
\newcommand{\sandbox}{initscript\xspace}
\newcommand{\sandboxes}{initscripts\xspace}
\newcommand{\Sandboxes}{Initscripts\xspace}

\newcommand{\exitnospace}{\cc{Exit}}
\newcommand{\exit}{\exitnospace\xspace}

\newcommand{\connectnospace}{\cc{Connect}}
\newcommand{\connect}{\connectnospace\xspace}
\newcommand{\sendnospace}{\cc{SendRPC}}
\newcommand{\recvnospace}{\cc{RecvRPCReply}}
\newcommand{\send}{\sendnospace\xspace}
\newcommand{\recv}{\recvnospace\xspace}
\newcommand{\txmsgnospace}{\cc{SendMsg}}
\newcommand{\txmsg}{\txmsgnospace\xspace}
\newcommand{\rxmsgnospace}{\cc{RecvMsg}}
\newcommand{\rxmsg}{\rxmsgnospace\xspace}

\newcommand{\rpcproxy}{\cc{WASMEngine}\xspace}

\newcommand{\rxconnnospace}{\cc{GetConn}}
\newcommand{\rxconn}{\rxconnnospace\xspace}
\newcommand{\txconnnospace}{\cc{TransferConn}}
\newcommand{\txconn}{\txconnnospace\xspace}

\newcommand{\proc}{\cc{proc}\xspace}
\newcommand{\procs}{\cc{proc}s\xspace}

\newcommand{\etcd}{\cc{etcd}\xspace}
\newcommand{\memcached}{\cc{memcached}\xspace}
\newcommand{\vecdb}{\cc{vecdb}\xspace}
\newcommand{\cached}{\cc{cached}\xspace}
\newcommand{\cacheds}{\cc{cached}s\xspace}

\newcommand{\imgrec}{\cc{imgrec}\xspace}
\newcommand{\imgrecw}{\cc{imgrec-wasm}\xspace}
\newcommand{\imgrecp}{\cc{imgrec-py}\xspace}

\newcommand{\vecdbspeedup}{1.72}

\newcommand{\cachedspeedup}{1.75}

\newcommand{\spicespeedup}{1.48}
\newcommand{\spicedl}{60}
\newcommand{\spiceweightsdl}{36}
\newcommand{\spicesetup}{50}

\newcommand{\sebsavgspeedup}{1.67}

\newcommand{\imgrecavgspeedup}{1.81}

\newcommand{\memcachedspeedup}{1.56}
\newcommand{\memcachedsnapsz}{200MB\xspace}

\newcommand{\etcdspeedup}{1.7}

\newcommand{\initscriptsize}{100KB\xspace}

\newcommand{\chalone}{challenge 1\xspace}

\newcommand{\chaltwo}{challenge 2\xspace}

\newcommand{\chalthree}{challenge 3\xspace}

\input{code/fmt}

\begin{document}


\date{\vspace{-\baselineskip}}

\title{Accelerating cloud application cold-start with \sandboxes}

\def\andnewline{\end{tabular} \\ \begin{tabular}[t]{c}}


\author{
Ariel Szekely, Hannah Gross, Robert T. Morris, Frans Kaashoek \\
MIT
}


\date{\vspace{-\baselineskip}}

\maketitle

\begin{abstract}
Serverless functions are a popular way of deploying cloud applications.
Because many of these functions are short-running and experience frequent
cold-starts, start latencies often dominate their execution latency.  Start
latency can be broken down into two components: \textit{setup} and
\textit{intitialization}.  Setup involves steps the cloud platform takes when
starting an application, such as downloading its binary and creating an
isolated execution environment. Initialization involves steps the application
takes after it has started running but before it can do useful work, such as
connecting to other services, coordinating to claim work, and downloading
inputs.

This paper contributes \textit{\sandboxes}, which provide a scriptable interface
for developers to specify their application's initialization routine to the
cloud platform. The platform can then run an application's \sandbox
and reduce application start latency by \textit{overlapping} setup and
initialization steps. Once the application is up and running, the \sandbox
bootstraps the application by transferring initialization results to it.

Using \sandboxes, we were able to speed up cold-starts of several of the
ServerlessBench~\cite{copik:sebs} Python applications by an average
\sebsavgspeedup$\times$ with no modifications to the application. \Sandboxes
also speed up start times for a serverless image recognition workload by
\imgrecavgspeedup$\times$, and off-the-shelf microservices like \etcd and
\memcached. 

\end{abstract}

\section{Introduction}
\label{s:intro}

The convenience and cost-effectiveness of serverless computing has driven
developers to refactor many cloud applications into serverless functions, and
to opt for a serverless architecture for new cloud applications.
Latency-sensitive user-facing websites written purely as serverless
applications are now
common~\cite{aws:serverless-websites,cloudflare:workers-sites,gcloud:faas}, and
new user-facing applications like Large Language Models (LLMs) and coding
agents often invoke serverless function tools and Model Context Protocol (MCP)
servers to construct their
responses~\cite{cloudflare:dynamic-workers,aws:ai-agents-on-serverless,gcloud:vertex-ai}.

Many of these serverless functions are short-running and experience frequent
cold-starts.  As a result, start latency often dominates their execution
latency~\cite{sharad:serverless,huawei-cold-start-trace,akkus:sand,du:catalyzer,li:rund,mitosis:wei}.
Traditionally, cold-start latency is defined as the time it takes a fresh
application instance to reach line 1 of \cc{main} on a new machine.  However,
applications must execute several additional steps before they can start doing
useful work. A more \textit{end-to-end} definition of cold-start latency, which
spans from the point an application is spawned until it begins processing work,
better captures the application developer's concerns.  Ultimately, this broader
definition of start latency more accurately reflects the serverless
infrastructure's overhead on the application's Service Level Objectives (SLOs).
The goal of this paper is to accelerate end-to-end cold-start latency.

\begin{figure}
  \scalebox{0.5}{\begin{tikzpicture}[
    divider/.style={thick, black},
    timeline/.style={-Stealth, very thick}
]

\def\setupW{5.0}
\def\initW{5.0}
\def\execW{1.4}

\def\setupX{0}
\pgfmathsetmacro{\initX}{\setupX + \setupW}
\pgfmathsetmacro{\execX}{\initX + \initW}
\pgfmathsetmacro{\totalW}{\execX + \execW}

\fill[blue!15]   (\setupX, -0.5) rectangle (\initX,  0.5);
\fill[orange!20] (\initX,  -0.5) rectangle (\execX,  0.5);
\fill[green!15]  (\execX,  -0.5) rectangle (\totalW, 0.5);

\draw[thick] (\setupX, -0.5) rectangle (\totalW, 0.5);
\draw[divider] (\initX, -0.5) -- (\initX, 0.5);
\draw[divider] (\execX, -0.5) -- (\execX, 0.5);

\node at (\setupX + \setupW/2, 0) {\textbf{Setup}};
\node at (\initX  + \initW/2,  0) {\textbf{Init}};
\node at (\execX  + \execW/2,  0) {\textbf{Exec}};

\node[left, font=\Large\bfseries] at (-0.4, 0) {a)};
\draw[timeline] (\setupX - 0.2, -0.65) -- (\totalW + 2.0, -0.65);
\node[right, font=\small] at (\totalW + 2.05, -0.65) {time};
\draw[thick] (0,      -0.55) -- (0,      -0.75);
\draw[thick] (\execX, -0.55) -- (\execX, -0.75);
\draw[thick, red] (0, -0.88) -- (0, -1.08) -- (\execX, -1.08) -- (\execX, -0.88);
\node[below, font=\normalsize\bfseries, red] at (\execX/2, -1.08) {Start Latency};

\def\initOff{0.3}
\pgfmathsetmacro{\parallelEnd}{\initOff + \initW}
\pgfmathsetmacro{\execEnd}{\parallelEnd + \execW}

\def\yTop{-2.0}       
\def\yMidTop{-3.0}    
\def\yMidBot{-3.0}    
\def\yBot{-4.0}       

\fill[blue!15]   (0,        \yMidTop) rectangle (\setupW,      \yTop);
\draw[thick]     (0,        \yMidTop) rectangle (\setupW,      \yTop);

\fill[orange!20] (\initOff, \yBot)    rectangle (\parallelEnd, \yMidBot);
\draw[thick]     (\initOff, \yBot)    rectangle (\parallelEnd, \yMidBot);

\pgfmathsetmacro{\execMid}{(\yTop + \yBot) / 2}
\fill[green!15]  (\parallelEnd, \yBot) rectangle (\execEnd, \yTop);
\draw[thick]     (\parallelEnd, \yBot) rectangle (\execEnd, \yTop);

\pgfmathsetmacro{\setupMid}{(\yTop    + \yMidTop) / 2}
\pgfmathsetmacro{\initMid} {(\yMidBot + \yBot)    / 2}

\node at (\setupW/2,               \setupMid) {\textbf{Setup}};
\node at (\initOff + \initW/2,     \initMid)  {\textbf{Init}};
\node at (\parallelEnd + \execW/2, \execMid)  {\textbf{Exec}};

\pgfmathsetmacro{\arrowY}{\yBot - 0.15}
\pgfmathsetmacro{\arrowEnd}{\execEnd + 2.0}
\node[left, font=\Large\bfseries] at (-0.4, -3.0) {b)};
\draw[timeline] (-0.2, \arrowY) -- (\arrowEnd, \arrowY);
\node[right, font=\small] at (\arrowEnd + 0.05, \arrowY) {time};
\pgfmathsetmacro{\tickTop}{\arrowY + 0.1}
\pgfmathsetmacro{\tickBot}{\arrowY - 0.1}
\draw[thick] (0,            \tickTop) -- (0,            \tickBot);
\draw[thick] (\parallelEnd, \tickTop) -- (\parallelEnd, \tickBot);
\pgfmathsetmacro{\bracketStart}{\tickBot - 0.13}
\pgfmathsetmacro{\bracketY}{\tickBot - 0.33}
\draw[thick, red] (0, \bracketStart) -- (0, \bracketY) -- (\parallelEnd, \bracketY) -- (\parallelEnd, \bracketStart);
\pgfmathsetmacro{\labelX}{\parallelEnd / 2}
\node[below, font=\normalsize\bfseries, red] at (\labelX, \bracketY) {Start Latency};

\end{tikzpicture}}
  \caption{Cloud application start timeline, \textbf{a)} without \sandboxes,
  and \textbf{b)} with \sandboxes.}
\label{fig:timeline}
\end{figure}

Start latency can be broken down into two components which we call
\textit{setup} and \textit{initialization}, shown in part \textbf{a)} of
\autoref{fig:timeline}.  Setup involves steps the platform takes before the
application starts running.  Setup is application-agnostic.  Examples include
downloading the application binary, creating an isolated execution environment,
and starting the application runtime. Initialization involves steps the
application takes after it is running, but which must complete before it can do
useful work.  Initialization is application-specific.  For example, a
serverless function may connect to a message queueing service to claim a task
to process, or a new microservice instance may query a load-balancer for its
shard assignment and download its shards from remote storage. Real-world traces
indicate that setup~\cite{sigmaos,mitosis:wei,huawei-cold-start-trace} and
initialization~\cite{fork-in-the-road,wei:krcore} are comparable in length for
many real-world applications: each takes O(100ms).

Prior work on reducing cloud application start time mostly focuses on reducing
setup costs.  Work on fast setup centers around lightweight isolation,
specialized hardware and kernel modifications for fast binary downloads, and
forking or snapshot-restore
mechanisms~\cite{sigmaos,jia:nightcore,shillaker:faasm,agache:firecracker,aws:step,kuchler:dandelion,mitosis:wei,fork-in-the-road}.
Some work has explored reducing initialization
costs~\cite{kuchler:dandelion,srivatsan:arca,deng:fix}. These systems push some
simple operations into the platform with APIs that allow developers to
statically declare application inputs and allow the platform to prefetch state
on behalf of the application.  Pushing initialization work into the platform in
this way reduces initialization latency, but prior approaches are
insufficiently expressive to capture the full gamut of application
initialization behavior.  For example, APIs which require developers to
statically declare function inputs do not enable input prefetching when the
input can only be determined at runtime, or fetching the input requires
exchanging RPCs with specialized storage services.

This paper proposes a missing programmatic interface for developers to specify
application initialization to the cloud platform: \textit{\sandboxes}.
\Sandboxes allow developers to write initialization programs which the platform
can run on their behalf.  Platforms can run \sandboxes in
parallel with the setup phase to \textit{overlap} setup and initialization and
reduce end-to-end start latency, as shown \autoref{fig:timeline} part
\textbf{b)}.  Once setup completes, the \sandbox bootstraps the application by
transferring initialization results to it.  \Sandboxes provide developers with
a scriptable interface which allows developers to express a wide range of
application initialization steps.  For example, \sandboxes can establish
connections to other services, issue RPCs and fetch state which can only be
identified at runtime. 

\Sandboxes must start with low latency in order to maximize the amount of
overlap with setup. A key design challenge (\chalone) in fast \sandbox
start is keeping \sandboxes small.  Small \sandboxes can be downloaded quickly
onto the new application instance's machine, and can overlap with a larger
fraction of the application container's setup to do more useful work.  Another
challenge (\chaltwo) is efficiently transferring initialization results
to the application.  The \sandbox needs to transfer established network
connections to the application.  Additionally, applications which load a
significant amount of state need to access the state with low overhead.  A
final challenge (\chalthree) is designing the \sandbox API to make it easy
to speed up start times for existing cloud applications with few modifications.

\Sandboxes are implemented as WASM modules, allowing the platform to run them
with strong isolation.  To resolve \chalone, \sandbox binaries are kept small,
around \initscriptsize, because the \sandbox host API is carefully designed to
implement a small set of primitives which can be composed to run common
initialization steps, namely establishing network connections and issuing RPCs.
This high-level RPC interface allows the platform to implement an asynchronous
RPC stack for the \sandbox and reduces \sandbox size.

To resolve \chaltwo, the \sandbox API is designed to enable efficient transfer
of initialization results to the application with a socket transfer and
message-passing API.  Application developers use this API to hand off
established connections to the application, and to move data between the
\sandbox and application container. The API's design enables applications to
claim and use connections established by the \sandbox, and to read
initialization state via shared memory.

Finally, to resolve \chalthree, the \sandbox API is designed to
support off-the-shelf cloud applications with few-to-no modifications. RPC
client libraries can support \sandbox-accelerated starts transparently to the
applications which depend on them. Using this approach, we used \sandboxes to
speed up start latency of several of the ServerlessBench~\cite{copik:sebs}
Python functions without any application code changes, and accelerate startup
of two popular open-source microservices, \memcached and \etcd, with a thin
compatibility layer.

We implement support for \sandboxes on \sigmaos~\cite{sigmaos}, a recent
multi-tenant cloud Operating System which supports fast container creation and
a uniform interface for a range of serverless and microservice tasks.

We use \sandboxes to accelerate cold-start latency of serveral unmodified
serverless applications from ServerlessBench~\cite{copik:sebs} by an average
\sebsavgspeedup$\times$, a WASM-based image recognition serverless function
\imgrecw and its Python analogue \imgrecp by \imgrecavgspeedup$\times$, two
unmodified real-world microservices, \memcached and \etcd, and two
representative custom-built microservices, \cached and \vecdb. \Sandboxes
cold-start onto new machines quickly, in 1-2ms including the cost of
downloading and isolating the \sandbox.  We also demonstrate that \sandboxes
are complementary to existing techniques which reduce setup and initialization
latency: \sandboxes speed up start latency for serverless functions which run
on a state-of-the-art snapshot-restore system~\cite{holmes:spice} by
\spicespeedup$\times$.

The main contributions of this work are:

\begin{itemize}
  
  \item \Sandboxes, a new interface for application developers to
  specify application initialization to cloud platforms, enabling overlap of
  setup and initialization (\autoref{s:design}).

  \item The design of the \sandbox API, which keeps \sandboxes small and
  enables efficient transfer of initialization results to applications with
  few-to-no modifications (\autoref{s:design}).

  \item An implementation of \sandboxes in \sigmaos (\autoref{s:impl}).
  
  \item An evaluation of \sandboxes and how they reduce start latency for
  off-the-shelf and custom cloud applications, and in conjunction with
  state-of-the-art cold-start speedup techniques (\autoref{s:eval}).

\end{itemize}

\section{Motivation}
\label{s:motiv}

In order to concretize the steps involved in setup and initialization we
examine \imgrec, a serverless image recognition application similar to the
image recognition application in ServerlessBench~\cite{copik:sebs}.  We compare
a WASM implementation (\imgrecw) and a Python implementation (\imgrecp) to
understand how the application runtime and isolation affects setup costs.

In order to initialize, \imgrecw and \imgrecp connect to Amazon's Simple Queue
Service (SQS)~\cite{aws:sqs}, a durable message queue built for serverless
applications, and claim a task.  They then download their model weights and the
task's input, perform inference, and store the result in S3.
\autoref{tab:setup-init-motiv-breakdown} shows the order-of-magintude cost of
each setup and initialization step of \imgrecw and \imgrecp.

\begin{table}[!h]
\centering
  \begin{tabular}{clll}
  \toprule
  Phase & Step & WASM & Python \\
  \midrule
  \multirow{3}{*}{Setup} & Binary download       & O(100ms) & O(1ms) \\
                         & Isolation             & O(1ms)   & O(100ms) \\
                         & Lang runtime init     & O(1ms)   & O(100ms) \\
                         & Fetch Dependencies    & \multicolumn{1}{c}{---} & O(100ms) \\
  \midrule
  \multirow{5}{*}{Init} & SQS GetTask RPC      & O(10ms) & O(10ms) \\
                        & Connect to S3        & O(1ms)  & O(1ms) \\
                        & Download inputs      & O(100ms) & O(100ms) \\
                        & Load state & O(10ms) & O(10ms) \\
  \bottomrule
  \end{tabular}
  \caption{Setup and initialization steps of an Image Recognition serverless
  application \imgrec, when implemented in WASM (\imgrecw) and Python
  (\imgrecp).}
  \label{tab:setup-init-motiv-breakdown}
\end{table}

\paragraph{Problem.} Modern cloud applications combine initialization with the
remainder of the application in a single package. As such, cloud platforms must
run application setup and initialization sequentially: initialization can only
begin once the full application has started running. During cold-starts,
unavoidable steps like binary downloads and starting the application runtime
push back the initialization phase by hundreds of milliseconds.

\paragraph{Goal.} Our goal it so separate initialization from the
application so that the platform can overlap setup and initialization,
and transfer the initialization to the application once it is has
started running.  In order to enable separation of initialization from the
rest of the application, the platform must provide a
\textit{scriptable} API which is sufficiently expressive to
capture a common set of cloud application initialization steps like
establishing connections and executing RPCs. For example, in
\imgrec later RPCs (e.g., the input download) depend on the results of earlier
ones (e.g., the claimed task). Furthermore, \imgrec knows which connections
to establish only at runtime: a new \imgrec instance only discovers the
ouptut destination once it has successfully claimed a task from SQS. The
following section describes how we achieve this goal using \sandboxes.

\section{Design}
\label{s:design}

This section describes the design of \sandboxes and their API. \Sandboxes are
WASM modules run by \rpcproxy, a per-machine daemon deployed by the cloud
platform (\autoref{ss:sandbox-wasm}).  Developers write \sandboxes with the
\sandbox API (\autoref{ss:sandbox-api}), which is designed to keep \sandboxes
small so they can be downloaded and start fast. The \sandbox result transfer
API (\autoref{ss:result-transfer}) enables \sandboxes to efficiently bootstrap
the application once it has started running. Its design enables existing
applications to use \sandboxes with few modifications.  This section also
discusses the \sandbox developer workflow (\autoref{ss:dev-workflow}), and
changes to APIs used to start cloud applications
(\autoref{ss:cloud-api-changes}).

\begin{figure}
  \scalebox{0.5}{\begin{tikzpicture}[
    node distance=2cm,
        component/.style={rectangle, rounded corners, draw=black, thick, minimum width=3cm, minimum height=2cm, align=center},
            process/.style={rectangle, draw=blue!60, thick, fill=blue!10, minimum width=2.5cm, minimum height=1.2cm, align=center},
                arrow/.style={-Stealth, thick},
                    message/.style={-Stealth, thick, dashed, red!70}
                    ]

\node[component, minimum width=5cm, minimum height=4.5cm, fill=blue!10] (rpcproxy) at (-3,1.5) {};
\node[above] at (rpcproxy.north) {\rpcproxy};

\node[component, minimum width=3cm, minimum height=3cm, fill=cyan!10] (s3) at (-10,2.3) {\textbf{S3}};

\node[rectangle, rounded corners, draw=orange!60, thick, fill=yellow!20, minimum width=1.5cm, minimum height=1cm, anchor=south west] (sandbox) at ($(rpcproxy.south west) + (0.3, 0.3)$) {\small\textbf{\Sandbox}};

\node[component, minimum width=4cm, minimum height=3.5cm, fill=purple!10] (container) at (3,1.5) {};
\node[above] at (container.north) {\textbf{Container}};

\node[rectangle, rounded corners, draw=purple!60, thick, fill=purple!20, minimum width=3cm, minimum height=2cm] (app) at (container.center) {\small\textbf{Application}};

\node[rectangle, draw=green!70, thick, fill=green!30, minimum width=0.8cm, minimum height=0.5cm] (microbox) at ($(app.south west) + (0.5, 0.35)$) {};

\node[rectangle, draw=gray!70, thick, fill=gray!5, minimum width=8cm, minimum height=2.2cm] (linux) at (0,-2.5) {};
\node[anchor=south west] at ($(linux.south west) + (0.2, 0.1)$) {\textbf{Linux}};

\node[rectangle, draw=black, thick, fill=green!15, minimum width=6cm, minimum height=0.8cm, anchor=north] (sharedmem) at ($(linux.north) + (0, -0.3)$) {\small\textbf{Shared Memory}};

\node[rectangle, draw=green!70, thick, fill=green!30, minimum width=1.5cm, minimum height=0.8cm, anchor=west] (darkbox) at (sharedmem.west) {};
\node[font=\small, anchor=south] at (darkbox.south) {\cc{buf_1}};

\coordinate (rpc_start) at ($(sandbox.north west) + (0.2, 0)$);
\draw[arrow] (rpc_start) -- ++(0,2.7) coordinate (rpc_point);
\node[font=\small, rotate=90, anchor=north west] at ($(rpc_start) + (0.1, 0)$) {\sendnospace\cc{()}};

\coordinate (u_start) at ($(sandbox.north east) + (-0.5, 0)$);
\coordinate (u_top) at ($(u_start) + (0, 2.5)$);
\coordinate (u_turn) at ($(u_top) + (0.3, 0)$);
\coordinate (u_end) at ($(sandbox.north east) + (-0.2, 0)$);
\draw[arrow] (u_start) -- (u_top) -- (u_turn) -- (u_end);
\node[font=\small, rotate=90, anchor=south west] at (u_start) {\recvnospace\cc{()}};
\node[font=\small, rotate=90, anchor=north] at ($(u_end) + (0, 1.25)$) {\cc{buf_1}};

\draw[arrow] (rpc_point) -- (rpc_point -| s3.east);
\node[above, font=\small] at ($(rpc_point -| s3.east) + (2, 0)$) {\cc{S3GetReq}};

\coordinate (s3_start) at ($(s3.east) + (0,0.3)$);
\coordinate (turn_down) at ($(rpcproxy.west) + (0.2, 0.3)$);
\coordinate (turn_right) at (turn_down |- sharedmem.center);
\draw[arrow] (s3_start) -- (s3_start -| turn_down) -- (turn_down) -- (turn_right) -- (sharedmem.west);
\node[below, font=\small] at ($(s3_start) + (2, 0)$) {\cc{S3GetRep}};

\draw[arrow] (sandbox.east) -- ++(2.3, 0) coordinate (sendmsg_point);
\node[above, font=\small] at ($(sandbox.east) + (1.15, 0)$) {\txmsgnospace\cc{(buf_1)}};

\draw[arrow] (sendmsg_point) -- (sendmsg_point |- app.west) -- (app.west);

\fill[green!15] (microbox.south west) -- (darkbox.north west) -- (darkbox.north east) -- (microbox.south east) -- cycle;

\draw[gray!60, thick] (microbox.south west) -- (darkbox.north west);
\draw[gray!60, thick] (microbox.south east) -- (darkbox.north east);

\end{tikzpicture}}
  \caption{A \sandbox fetching initialization state and communicating it to its
  application via the \rpcproxy. The \sandbox sends an RPC requesting the
  application's initialization state to S3 via \rpcproxy using \send, and
  blocks until the reply comes back using \recv. \rpcproxy writes the response
  from S3 directly into a shared memory region, and sends the buffer \cc{buf_1}
  containing the reply to the \sandbox. The \sandbox forwards \cc{buf_1} to the
  application, which fetches the state by mapping the shared memory region and
  reading \cc{buf_1} directly.}
\label{fig:sys-diagram}
\end{figure}

\subsection{\Sandboxes: scriptable WASM init routines}
\label{ss:sandbox-wasm}

Multi-tenant cloud platforms that support \sandboxes need to start them quickly
and with strong isolation. In order to do so, \sandboxes are implemented as
WASM modules executed by \rpcproxy, a per-machine daemon run by the platform.
Because \sandboxes are WASM modules, they don't have access to a full Linux
environment with standard syscalls. Instead, they use the \sandbox API to
communicate with the outside world and carry out initialization steps.
\autoref{fig:sys-diagram} illustrates how a \sandbox uses the \sandbox API to
interact with \rpcproxy and fetch initialization state from
S3. \fk{inline the caption into the text? it makes co-sandboxes concrete
and helps the reader understand what the plan is} The following
section describes the design of the \sandbox API.

\subsection{\Sandbox API}
\label{ss:sandbox-api}

\begin{table*}[t]
  \begin{center}
    \renewcommand{\arraystretch}{1.1}
\begin{tabularx}{\linewidth}{p{6.8cm}L}
  \toprule
  Methods & Description \\
  \midrule
  \connectnospace\cc{(addr)} $\rightarrow$ \cc{conn_fd} & Establish network connection to destination \cc{addr} (IP or DNS name)  \\
  \sendnospace\cc{(conn_fd, rpc_id, buf)} & Asynchronously send \cc{rpc_id}'s \cc{buf} over \cc{conn_fd} \\
  \recvnospace\cc{(conn_fd, rpc_id)} $\rightarrow$ \cc{buf} & Block until \cc{rpc_id} is complete and return its reply \cc{buf} \\
  \exitnospace\cc{(status)} & Exit with \cc{status} \\
 \bottomrule
\end{tabularx}
  \caption{API developers use to write \sandboxes.}
\label{tab:sandbox-api}
\end{center}
\end{table*}

The primary design goal of the \sandbox API, shown in
\autoref{tab:sandbox-api}, is to keep \sandboxes small (\chalone) by shifting
as much functionality as possible into the platform.  In service of this goal,
the \sandbox API is high-level and provides a small set of functions for
connection establishment and RPCs: \connect, \send, and \recv. The high-level
design of the API reduces the compiled binary size of \sandboxes because
\rpcproxy can supply an implementation of functionality needed to interact with
cloud services such as client-side RPC and networking stacks, support for
long-lived sessions, and authentication.  

\connect establishes a network connection to a service with DNS name or IP
address \cc{addr}, and returns a file descriptor \cc{conn_fd}.  \send
asynchronously dispatches a marshaled RPC on the connection named by
\cc{conn_fd}. \recv blocks until a response is received, and returns a buffer
containing the response to the \sandbox.

Asynchronous \send is an important feature of the RPC API since WASM modules
are single-threaded. \send allows the \sandbox to extract parallelism by
delegating asynchronous communication to the host-side of the RPC API. This
allows the \sandbox to, for example, construct and send other RPCs while the
host runs RPCs on its behalf.

\exit terminates the \sandbox and reports the exit \cc{status} to the platform.

\subsection{\Sandbox result transfer API}
\label{ss:result-transfer}

\begin{table*}[t]
  \begin{center}
    \renewcommand{\arraystretch}{1.1}
\begin{tabularx}{\linewidth}{p{6.5cm}L}
  \toprule
  Methods & Description \\
  \midrule
  \txmsgnospace\cc{(msg_id, buf)} & Send \cc{buf} between \sandbox/container \\
  \rxmsgnospace\cc{(msg_id)} $\rightarrow$ \cc{buf} & Receive a \cc{buf} message from \sandbox/container \\
  \txconnnospace\cc{(conn_id, addr, conn_fd)} & Transfer ownership of cached connection to \sandbox/container \\
  \rxconnnospace\cc{(conn_id, addr)} $\rightarrow$ \cc{conn_fd} & Receive ownership of connection from \sandbox/container \\
 \bottomrule
\end{tabularx}
	\caption{\Sandbox result transfer API that developers use to bootstrap a
	\sandbox's applications once the application is up and running.}
\label{tab:sandbox-comm-api}
\end{center}
\end{table*}

The \sandbox result transfer API, shown in \autoref{tab:sandbox-comm-api}, is
designed to efficiently pass initialization results from the \sandbox to the
application (\chaltwo), and to make it possible for existing applications to use
\sandboxes with few modifications (\chalthree).

\txmsg and \rxmsg exchange buffers of data, such as application state or
serverless inputs downloaded by the \sandbox.  \rpcproxy uses shared memory to
enable zero-copy transfer of data from the \sandbox to the application, an
important feature for applications that load a significant amount of state.

\rpcproxy sets up a shared memory region to store \sandbox's RPC results when a
new application instance is assigned to a machine.  \rpcproxy dispatches the
\sandbox's RPCs and writes the replies directly into the shared memory region.
When the \sandbox calls \recv, \rpcproxy returns a buffer backed by this shared
memory region to allow the \sandbox to access the reply without copying it. The
\sandbox passes the buffer on to the application using \txmsg.

After the application starts, its \sandbox library maps the
shared memory region \rpcproxy set up for it. The application calls \rxmsg to
receive buffers passed to it by the \sandbox, and \rpcproxy returns
corresponding buffers backed by the shared memory region. \rpcproxy
synchronizes access to RPC results, because long-running RPCs initiated by the
\sandbox may not complete until after the application has started running and
called \rxmsg.

\txconn and \rxconn transfer ownership of a network connection, allowing the
application to use the connection directly. By claiming ownership of a
connection from its \sandbox, the application avoids re-establishing and
re-authenticating the connection if it needs to issue follow-up RPCs to the
remote service.  \rpcproxy implements \txconn and \rxconn by passing the
connection's Linux file-descriptor to the application over a socket.

The result transfer API employs a flexible naming system in which developers
define the identifiers for connections and messages. RPC client library
developers can use this naming system to support \sandbox-assisted
initialization below the RPC library's API and transparently to the
application. For example, the RPC library can identify RPCs with a counter or
hash of API function arguments.

The application developer then follows the naming scheme when writing their
\sandbox. For example, the \sandbox may use a counter to line up the RPCs it
issues with the order of initialization RPCs in the application.  When the
application starts up and tries to issue its initialization RPCs, the RPC
client library intercepts the RPCs and serves the RPC results from the \sandbox
via \rpcproxy.  \rpcproxy uses the message and connection IDs to synchronize
access to RPC results and ensure both the \sandbox and application see a
consistent transcript of initialization RPCs and results.  \rpcproxy prevents
the application from getting ahead of the \sandbox and issuing RPCs that the
\sandbox has yet to send or re-issuing RPCs that are already in progress.

\subsection{Developer workflow}
\label{ss:dev-workflow}

Developers can write \sandboxes in any language that compiles to WASM.
Developers upload their \sandbox to the cloud provider separately from the
application binary and declare resource reservations for it as they do for
application containers and serverless functions today. 

Developers can optionally specify a byte slice as an input argument to a
\sandbox. This can be used, for example, to communicate function inputs or
credentials for the \sandbox to use when sending RPCs and establishing sessions
to other services.

\subsection{Cloud API changes}
\label{ss:cloud-api-changes}

Cloud providers can
support \sandboxes by augmenting existing APIs. For example, application 
launch and serverless function invocation APIs that specify a function or
application container ID can be augmented to specify a \sandbox to run.

\section{Implementation}
\label{s:impl}

We implemented \sandboxes on \sigmaos~\cite{sigmaos}, a recent cloud operating
system built on Linux which supports applications with fast container creation
times. The remainder of this section discusses the details of the new \sigmaos
components which implement the \sandbox API (\autoref{ss:rpcproxy}),
modifications to \sigmaos APIs that developers use to spawn tasks
(\autoref{ss:sigmaos-api-mods}), and implementation details of the \sandboxes
we wrote to support our applications and evaluation
(\autoref{ss:sandbox-impl}).

\subsection{\rpcproxy}
\label{ss:rpcproxy}

The \rpcproxy implements the \sandbox API (\autoref{tab:sandbox-api}) and makes
the results of \sandbox RPCs available to new application instances via the
\sandbox result transfer API (\autoref{tab:sandbox-comm-api}). \rpcproxy runs
as a per-machine daemon on every node managed by the cloud platform. It
collaborates with the platform to manage the application instance lifecycle.
\rpcproxy is written in Go and runs \sandboxes using the Wasmer WebAssembly
runtime~\cite{wasmer}.

When a new application instance starts, \rpcproxy quickly downloads its
\sandbox and maps a region of shared memory using the Linux POSIX shared-memory
APIs to store RPC replies. It then runs the \sandbox in a WASM sandbox,
establishing and caching connections to downstream services, and forwarding
marshaled RPCs to them as requested by the \sandbox.

\rpcproxy communicates with an application instance via a Unix named pipe
mounted into the container's filesystem in order to coordinate access to
the initialization results sent by the \sandbox. \rpcproxy implements a simple
protocol over the pipe to block microservice \rxmsg requests in the event
until the \sandbox sends the corresponding \txmsg, and vice versa. In the event
that the \sandbox or application crashes with an error, \rxmsg will return
an error to the caller.

When it starts up, the application maps the shared memory region created by
\rpcproxy using POSIX shared-memory APIs. \txmsg and \rxmsg exchange pointers
into this region, allowing the application to directly read data passed to it
by the \sandbox.

\subsection{\sigmaos API modifications}
\label{ss:sigmaos-api-mods}

We augmented the \sigmaos Spawn API include a serialized \sandbox binary for
the \proc and an input byte slice for the \sandbox. Additionally, we added
support for resource reservations for \sandboxes.

In order to support unmodified Linux microservices like \memcached and \etcd,
we added a GVisor isolation backend for \sigmaos procs. We also added support
for Python applications with dynamic imports in order to support the
ServerlessBench applications and \imgrecp, and a new isolation backend to run
WASM \procs like \imgrecw. For the comparison to Spice~\cite{holmes:spice}, we
added Junction~\cite{fried:junction} as an isolation backend as well.

Additionally, we added support for WASM \procs and
applications isolated by Spice~\cite{holmes:spice} in order to run \imgrecw and
the snapshot-restore benchmarks.

\subsection{\Sandbox implementation details}
\label{ss:sandbox-impl}

We wrote \sandboxes for the applications in Rust. \sandboxes import a
library which declares external functions corresponding to the \sandbox API
(\autoref{tab:sandbox-api}) and \sandbox result transfer API
(\autoref{tab:sandbox-comm-api}). 

\section{Case-studies: applications using \sandboxes}
\label{s:case-studies}

We ported several off-the-shelf cloud applications to use \sandboxes, including
all of the ServerlessBench~\cite{copik:sebs} Python applications and two
microservice applications, \etcd and \memcached.  We also wrote some
applications from scratch, including serverless applications like \imgrecw and
\imgrecp, as well as microservices like \vecdb, an in-memory vector
database, and \cached, a \sandbox-native analogue to \memcached. The remainder
of this section discusses the experience of adapting existing applications
(\autoref{ss:cs-existing}) to benefit from \sandbox-accelerated cold-starts and
writing new (\autoref{ss:cs-new}) \sandbox-accelerated applications from
scratch. 

\subsection{Fast cold-start for off-the-shelf applications}
\label{ss:cs-existing}

Developers can use \sandboxes to accelerate cold-start for existing
applications by writing a simple compatibility layer which communicates with
the \sandbox, and makes the fetched state available to the application
(\autoref{ss:result-transfer}).

\begin{figure}[t]
  \input{code/sebs-storage-api.tex}
  \caption{ServerlessBench Python storage API.}
  \label{fig:sebs-storage-api}
\end{figure}

\begin{figure}[t]
  \input{code/sebs-storage-impl.tex}
  \caption{ServerlessBench storage client library modified to use \sandboxes
  for initialization results. The library developer defines the initialization
  result naming scheme as a simple counter so that the library can fetch
  results from the \sandbox.}
  \label{fig:sebs-storage-impl}
\end{figure}

For example, ServerlessBench's Python applications use a
\cc{get}/\cc{put}-style API typical of client libraries used to access cloud
storage services like S3, shown in \autoref{fig:sebs-storage-api}. We added
\sandbox support to all the ServerlessBench Python functions with no
modifications to the applications themselves. Instead, we changed the client
library implementation to intercept the initialization RPCs and communicate
with the \sandbox rather than send the RPCs to the remote storage service. The
\sandbox passes the stored results of the initialization RPCs back to the
client library, which returns the results to the application.

\begin{figure}[t]
  \input{code/sebs-imgrec-app.tex}
  \caption{The unmodified ServerlessBench ImageRecognition application benefits
  from \sandbox support in the storage client library.}
  \label{fig:sebs-imgrec-app}
\end{figure}

\begin{figure}[t]
  \input{code/sebs-storage-cosandbox.tex}
  \caption{The ImageRecognition application's \sandbox uses the storage
  library-defined result naming scheme to fetch and pass on the application's
  initialization state.}
  \label{fig:sebs-storage-cosandbox}
\end{figure}

\autoref{fig:sebs-storage-impl} shows the implementation of the download
function in the storage library with \sandbox support. The \sandbox result
transfer API allows the storage library developer to choose a naming system to
communicate results: in this case, a counter. Developers who wish to use the
library and speed up their application's start time with a \sandbox can do so
by using the same naming system defined by the storage library developer.

For example, the ServerlessBench ImageRecognition application shown in
\autoref{fig:sebs-imgrec-app} first downloads its input image and then
downloads its model weights. The application developer can make use of \sandbox
support in the storage library by writing a \sandbox which fetches the
initialization state in the same order, as shown in
\autoref{fig:sebs-storage-cosandbox}. Since the application developer followed
the naming scheme defined by the storage library, the unmodified Image
Recognition will transparently benefit from \sandbox-accelerated start time.
The \sandbox will run while the platform carries out the application's setup,
and will bootstrap the application once it is up and running.

One reason \sandboxes are able to benefit existing applications with few-to-no
modifications is that modern cloud applications rely on RPC client libraries to
communicate with external services. These libraries already abstract many of
the details of this communication: applications are relatively oblivious to
connection establishment, network transport, marshalling and unmarshalling,
management and interpretation of the retrieve data. By modifying the RPC
libraries to take advantage of the \sandbox API's efficient initialization
result transfer and defining a clear naming scheme, library developers enable
applications which depend on those libraries to benefit as well.

\subsection{Fast cold-start for \sandbox-native applications}
\label{ss:cs-new}

Although \sandboxes can benefit off-the-shelf applications, \sandbox-native
application 
can achieve even greater cold-start latency reduction by using the \sandbox
API's support for shared memory and connection passing. These applications
can build their internal data structures from shared memory regions populated
by the \sandbox to avoid any \cc{memcopy}ing and can reuse connections
established by the \sandbox. Two examples are \vecdb and \cached.

\vecdb is a sharded in-memory vector database implemented in C++. Clients send
\vecdb a request including an input vector as well as a range of vectors to
search. \vecdb performs a cosine-similarity search over its vector database,
and returns the IDs of the vectors closest to the input vector.  \cached is an
in-memory sharded key-value cache implemented in C++, similar to
\cc{memcached}~\cite{memcached}.

\begin{figure}[t]
  \input{code/vecdb-initscript.tex}
  \caption{\vecdb \sandbox fetches the shard assignment and sends it to the
  microservice application.}
  \label{fig:vecdb-initscript-code}
\end{figure}

\begin{figure}[t]
  \input{code/vecdb-initscript-retrieval.tex}
  \caption{\vecdb microservice application receives state from its \sandbox.}
  \label{fig:vecdb-initscript-retrieval-code}
\end{figure}

\vecdb uses \sandboxes to accelerate scaling up. When a new \vecdb instance
starts, its \sandbox downloads its shard of the vector space, which may be
stored in an in-memory cache service (like \cached), in a database, or in
provider-managed storage (like S3).  \cached operates similarly to \vecdb by
using its \sandbox to fetch its shards from existing peer \cacheds.

\autoref{fig:vecdb-initscript-code} shows the implementation of the \sandbox
which fetches \vecdb's shards and sends them to the new \vecdb instance.
\autoref{fig:vecdb-initscript-retrieval-code} shows how the newly running
\vecdb communicates with its \sandbox to receive its state and load the state
into its internal data structures.

The message-passing API which the \sandbox and \vecdb use to communicate
initialization results, \txmsg and \rxmsg, are implemented using
shared-memory set up by \rpcproxy. This allows the \sandbox to issue RPCs to
fetch \vecdb's shards and pass the results back to \vecdb without additional
memcopies. \vecdb builds its internal data structures when loading its shards
by simply setting references to the shared memory region containing them.

\section{Evaluation}
\label{s:eval}

The goal of \sandboxes is to reduce end-to-end cold-start latency with few
changes to applications.  In order to determine whether \sandboxes meet their
goal, this evaluation answers the following questions:

\begin{enumerate}

	\item How much do \sandboxes speed up off-the-shelf applications' start latency?
	(\autoref{ss:eval-start-speedup-existing})

	\item How much more do \sandbox-native applications improve their start latency?
  (\autoref{ss:eval-start-speedup-new})

	\item How small can \sandboxes be? (\autoref{ss:eval-sandbox-size})

  \item Are \sandboxes able to speed up start latency on platforms with
  fast setup? (\autoref{ss:eval-blink})

\end{enumerate}

\paragraph{Experimental setup.} We evaluate \sandboxes using 8 AWS EC2
m6i.4xlarge VMs with 16 vCPUs, backed by 2.9GHz Intel Ice Lake 8375C CPUs, 64
GiB of memory, up to 12.5Gbps network bandwidth, and up to 10 Gbps EBS burst
bandwidth. The client program which spawns microservices and invokes serverless
functions runs on one machine, and communicates with a \sigmaos cluster running
on the other machines to run the experimental workloads. For the comparison
to state-of-the-art snapshot-restore systems, we evaluate on a bare-metal
server with an Intel Xeon Gold 5420+ CPU with 56 logical cores.

We evaluate several of the Python ServerlessBench~\cite{copik:sebs}
applications.  We also evaluate a serverless image recognition application
\imgrecw built in Rust and compiled to WASM, an equivalent version \imgrecp
written in Python, and 4 microservice applications, \etcd, \memcached, \vecdb,
and \cached.  We use GVisor containers to isolate \etcd, \memcached,
fast-starting \xcontainers to isolate \cached and \vecdb, the Wasmer runtime to
isolate \imgrecw, and \xcontainers to isolate the Python applications.

\subsection{\Sandboxes reduce off-the-shelf application start latency}
\label{ss:eval-start-speedup-existing}

One goal of the \sandbox design is to enable developers to reduce existing
applications' start latency with few application modifications. In order to
evaluate whether \sandboxes achieve this goal, we compare the start latency
without and with \sandboxes of several unmodified Python ServerlessBench
applications, \imgrecw and \imgrecp, and two popular microservice applications,
\etcd and \memcached.

We define start latency as the time from which a new application instance is
spawned, and the time at which it starts to handle the first request. The
following benchmarks report the start latency of each application during a
cold-start, defined as the first time a machine runs a given application. In
particular, this means that the costs of binary download and container creation
are included in the setup latency, and the costs of fetching inputs, model
weights, and soft-state are included in initialization latency.
A good result would show that \sandboxes decrease the start-latency of the
applications, because they overlap setup and initialization.

\begin{figure*}
 \includegraphics[width=\linewidth,keepaspectratio]{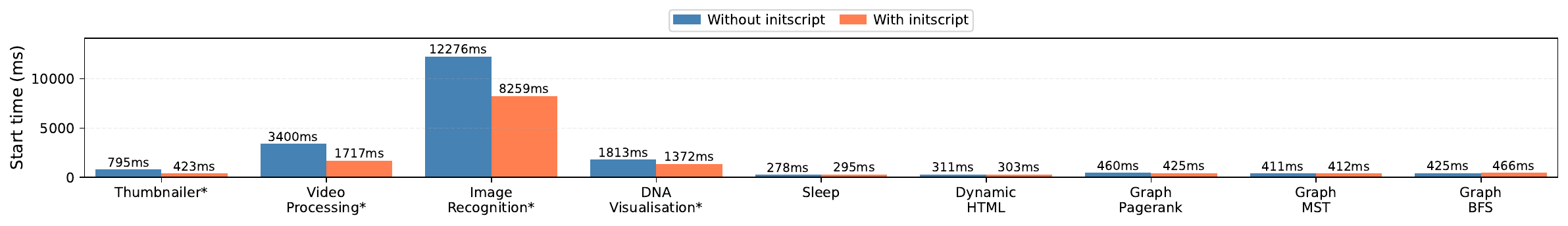}
  \caption{Start time of the Python ServerlessBench benchmarks in \sigmaos
  without and with \sandboxes.}
  \label{fig:sebs-start-latency}
\end{figure*}

\autoref{fig:sebs-start-latency} shows the start latency of several
ServerlessBench applications. Functions marked by a (\textbf{*}) download their
inputs and model weights from S3.  \Sandboxes speed up their start-latency by
an average of \sebsavgspeedup$\times$, because \sandboxes pre-establish
connections to S3 and fetch the inputs and model weights during setup. The
remaining serverless applications do not experience a reduction in
start-latency when run with \sandboxes because they are purely functional. They
do not fetch any input or write any output. Instead, they return their output
as a string response to the invocation request. Some of these functions exhibit
slightly higher start latency when running with \sandboxes due to variability
in download times of the packages they import. 

\begin{figure}
 \includegraphics[width=\linewidth,keepaspectratio]{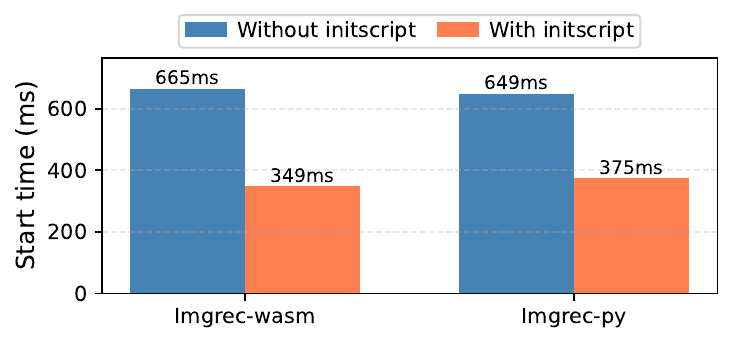}
  \caption{Start time of \imgrecw and \imgrecp in \sigmaos
  without and with \sandboxes.}
  \label{fig:start-latency-imgrec}
\end{figure}

\autoref{fig:start-latency-imgrec} shows the start latency of \imgrecw and
\imgrecp when running without and with \sandboxes.  Both implementations
benefit from \sandbox-accelerated start time without any code changes. In this
benchmark, each \imgrec application downloads its 13.3MB model weights and the
4.9MB input image from S3 during initialization. Together, the downloads take
312ms on average.  

\Sandboxes speed up the \imgrec applications' start time by
\imgrecavgspeedup$\times$ across \imgrecw and \imgrecp. Even though \imgrecw
benefits from fast WASM isolation, downloading its large (45.6MB) WASM binary
gives its \sandbox ample time to execute initialization RPCs and fetch its
input and model weights. Conversely, \imgrecp's binary is small (3.4KB) because
it runs with a platform-supplied Python interpreter. However, the cost of
starting an isolated Python interpreter and dynamically loading the function's
libraries also gives the \sandbox time to carry out initialization steps.


\begin{figure}
 \includegraphics[width=\linewidth,keepaspectratio]{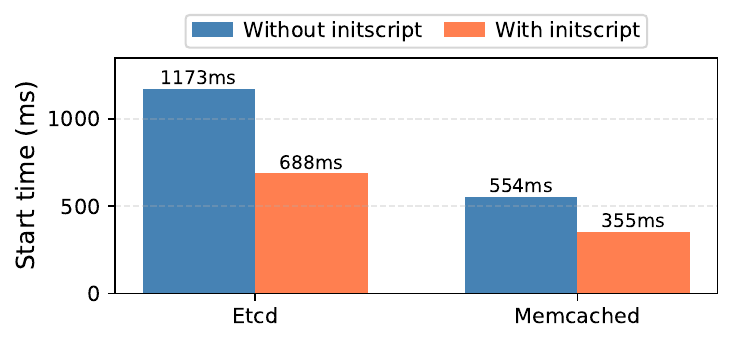}
  \caption{Start time of \etcd and \memcached in \sigmaos without and with
  \sandboxes.}
  \label{fig:start-latency-shims}
\end{figure}

\autoref{fig:start-latency-shims} shows the start latency of \etcd and
\memcached when running without and with \sandboxes. In this benchmark, \etcd
initializes by restoring a 14MB snapshot~\cite{etcd:recovery}, and \memcached
initializes with a warm restart~\cite{memcached:warm-restart} from a \memcachedsnapsz
snapshot. \etcd and \memcached fetch their state from a \sigmaos storage service.  \etcd starts
\etcdspeedup$\times$ faster, and \memcached starts \memcachedspeedup$\times$
faster when using \sandboxes. The start time speedup \sandboxes provide to
\etcd and \memcached is less than what they provide the previous serverless
examples, because initializing these microservices requires reconstructing
complex internal data structures from the downloaded state. Since \etcd's and
\memcached's unmodified application binaries are not set up to share memory
with the \sandbox, the \sandbox cannot carry out this initialization step for
them.

\begin{figure*}
 \includegraphics[width=\linewidth,keepaspectratio]{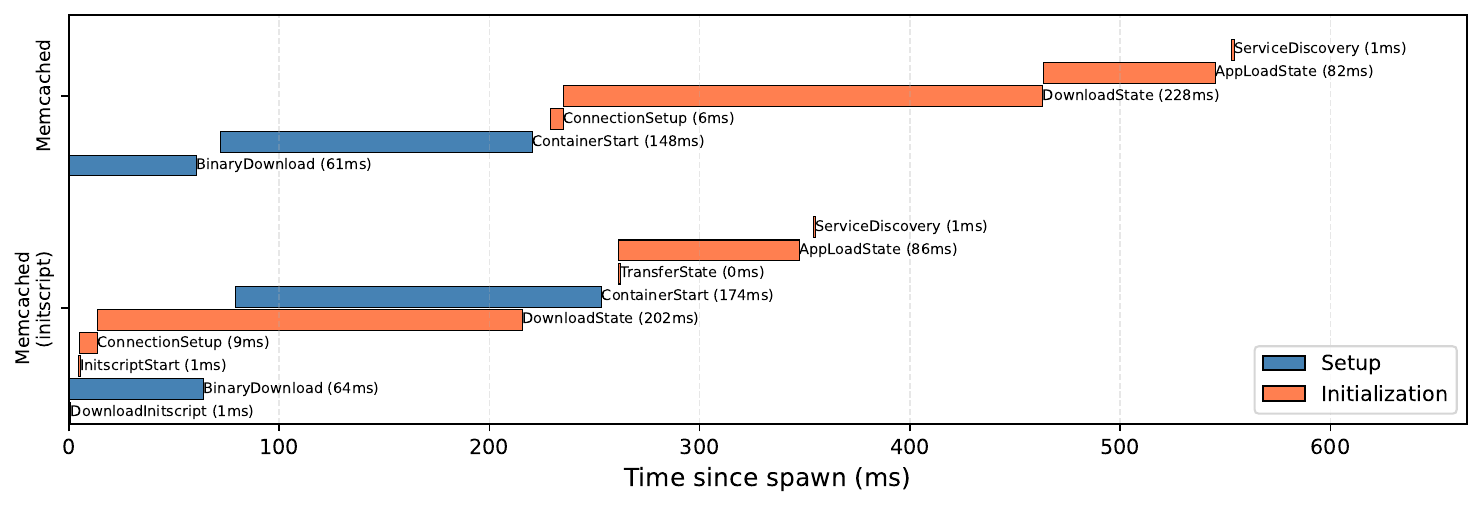}
  \caption{Start latency breakdown for \memcached without and with \sandboxes.}
  \label{fig:memcached-start-latency-breakdown}
\end{figure*}

In order to understand how \sandboxes speed up \memcached and other existing
applications, we break down \memcached's start latency without and with
\sandboxes in \autoref{fig:memcached-start-latency-breakdown}. The \sandbox
downloads the \memcached snapshot from the storage service and transfers it to a small 180 line
compatibility shim via the \sandbox communication API. The shim writes the
snapshot to the \memcached container's local filesystem, and directs \memcached
to it.  \memcached then uses its warm restart~\cite{memcached:warm-restart}
feature to parse the snapshot file and reconstruct its internal data
structures. The extra copies induced by writing the snapshot to the container
filesystem reduces the speedup \sandboxes provide to \memcached: the shim and
\memcached spend around 85ms exchanging state via the filesystem.

\paragraph{Summary.} \Sandboxes are able to reduce end-to-end start latency for
several existing serverless and microservice applications. Some applications
benefit from \sandboxes without any modifications, while others require modest
changes.

\subsection{\Sandbox-native applications achieve additional speedup with \sandboxes}
\label{ss:eval-start-speedup-new}

Another goal of the \sandbox API is to support efficient transfer of
initialization results to the application. The evaluation of \memcached and
\etcd in \autoref{ss:eval-start-speedup-existing} shows that unmodified
applications which load initialization state into complex internal data
structures pay a penalty in start latency. This section uses two
\sandbox-native microservice applications, \vecdb and \cached, to explore how
much of the lost start time can be recovered by restructuring applications
around the \sandbox result transfer API.

\vecdb and \cached are two representative sharded soft-state microservices.
In this benchmark, we cold-start a new \vecdb instance and a new \cached
instance on a new machine in the \sigmaos cluster.  \vecdb and \cached
initialize by downloading shards of their service's data assigned to them, which
amount to 9.2MB and 15MB respectively.  \vecdb fetches its state from an
in-memory key-value cache, and \cached fetches its state from existing peer
\cacheds.

\vecdb and \cached are written from scratch and designed to reconstruct their
internal data structures with the \sandbox result transfer API in mind.  Both
internally assign pointers to the memory region they share with their \sandbox,
enabling them to set up their state without copying their initialization data.
Due to their optimized use of the \sandbox result transfer API, a good result
would show \vecdb and \cached experiencing start latency acceleration closer to
that experienced by the serverless functions.

\begin{figure}
 \includegraphics[width=\linewidth,keepaspectratio]{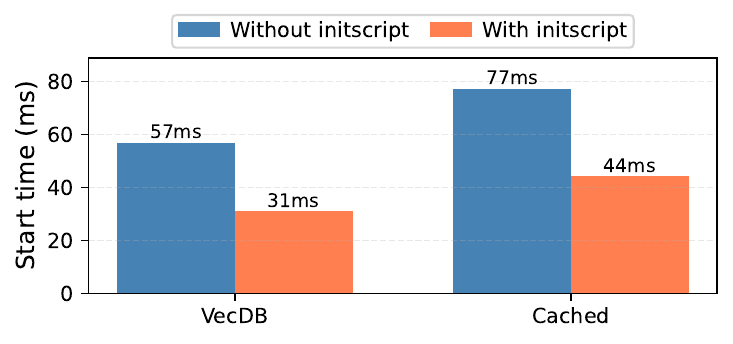}
  \caption{Start time of \vecdb and \cached in \sigmaos without and with
  \sandboxes.}
  \label{fig:start-latency-cpp}
\end{figure}

\autoref{fig:start-latency-cpp} shows the start latency of \vecdb and \cached
without and with \sandboxes.  \vecdb starts \vecdbspeedup$\times$ faster and
\cached starts \cachedspeedup$\times$ faster. \Sandboxes speed up both
applications' starts because the \sandbox fetchees the application's state in
parallel with the container setup, runtime initialization, and binary download.
\begin{figure*}
 \includegraphics[width=\linewidth,keepaspectratio]{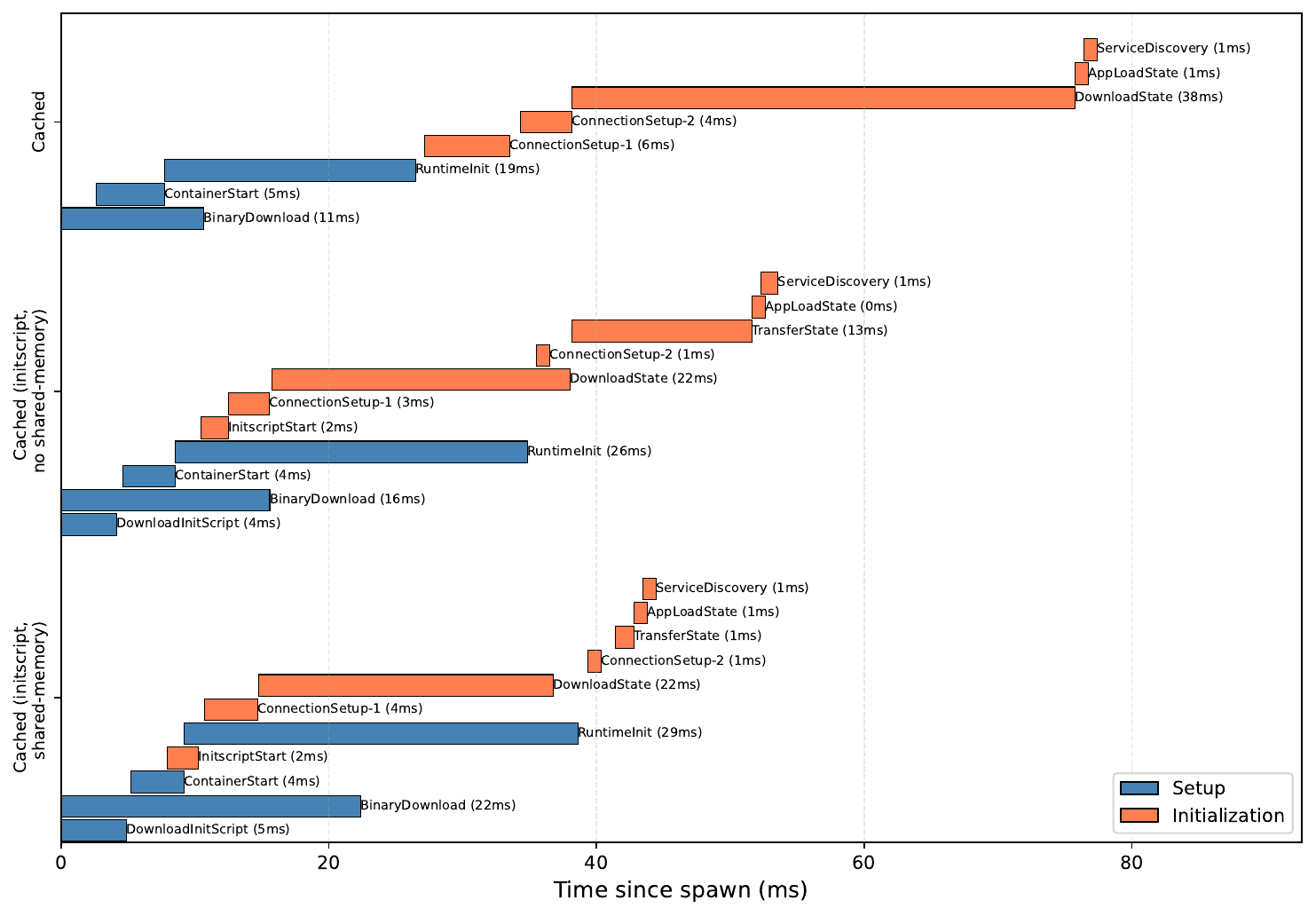}
  \caption{Start latency breakdown for \cached without and with \sandboxes,
  without and with shared-memory transfer to pass results from the \sandbox to
  \cached.}
  \label{fig:cached-start-latency-breakdown}
\end{figure*}

In order to understand why these \sandbox-native applications achieve better
speedup than \etcd and \memcached, we show a timeline which breaks down
\cached's cold-start latency in \autoref{fig:cached-start-latency-breakdown}.
The \textbf{\cached (\sandbox, no shared memory)} breakdown copies data when
loading the initialization state from the \sandbox into \cached's internal data
structures.  The additional memcopies slow down state transfer from the
\sandbox to \cached from 1 millisecond to 13 millseconds, eroding the speedup
benefit of \sandboxes by 39\%. This is concordant with the start latency
penalty experienced by \memcached and \etcd.  When utilizing the shared memory
features of the \sandbox result transfer API, however, \cached achieves much
more speedup.  This highlights the importance of efficiently transferring
results with the \sandbox result transfer API.

\paragraph{Summary.} Efficiency enabled by the \sandbox result transfer API
design allows applications to derive additional start latency speedup when
designed to use the API.  This is particularly important for applications which
reconstruct complex internal data structures during initialization.

\subsection{\Sandboxes can be small}
\label{ss:eval-sandbox-size}

One reason \sandboxes can start quickly is that \sandbox binaries can be
much smaller than application binaries, which makes downloading a \sandbox's
binary fast. \autoref{fig:cached-start-latency-breakdown} and
\autoref{fig:memcached-start-latency-breakdown} show that \cached and
\memcached's 200KB \sandbox binary downloads in 2-3ms, whereas downloading
their multi-MB binaries takes 10s to 100s of milliseconds.  Prior work shows
that real-world cloud application binaries and container images range from
several megabytes to hundreds of megabytes~\cite{faasnet:wang,brooker:lambda}
in size.  In the event of a cold-start, network bandwidth and binary download
latency become a limiting factor for performance. \Sandbox binaries can be much
smaller, on the order of a hundred kilobytes, because initialization is a small
fraction of a full application's functionality.

\begin{table}[t]
  \begin{center}
    \renewcommand{\arraystretch}{1.1}
\begin{tabularx}{\linewidth}{lR}
  \toprule
  Component & KB  \\
  \midrule
  Init RPC marshaling        &    13 \\
  Clnt RPC stubs             & 1,330 \\
  RPC stack                  &   469 \\
  Logging                    &    23 \\
  Perf monitoring            &    53 \\
  Srv RPC marshaling         &    23 \\
  Application impl           &   432 \\
  Exceptions                 &    54 \\
  Misc                       &    88 \\
  \midrule
  Total                      & 2,485 \\
 \bottomrule
\end{tabularx}
\end{center}
  \caption{Contributors to the size of a stripped \vecdb binary, as reported by
  Bloaty~\cite{bloaty}.}
\label{tab:us-sz-breakdown}
\end{table}

Take \vecdb, for example, a typical C++ microservice built on \sigmaos.
\autoref{tab:us-sz-breakdown} shows a breakdown of the components which
contribute to the size of the \vecdb binary, as reported by
Bloaty~\cite{bloaty}. \vecdb needs several client RPC stubs to operate,
telemetry and performance monitoring, a full server-side RPC stack and support
for marshaling client requests, and support for marshaling RPCs of several
services it depends on and the associated client RPC stubs for each.

\begin{table}[t]
  \begin{center}
    \renewcommand{\arraystretch}{1.1}
\begin{tabularx}{\linewidth}{lrr}
  \toprule
  Component & \rpcproxy KB & \Sandbox KB \\
  \midrule
  Init RPC marshaling        &       & 13 \\
  Clnt RPC stubs             & 1,330 &    \\
  RPC stack                  &   469 &    \\
  Logging                    &    23 &    \\
  Perf monitoring            &    53 &    \\
  Exceptions                 &       &  54 \\
  Misc                       &       &  88 \\
  \midrule
  Total                      &  1875 & 155\\
 \bottomrule
\end{tabularx}
\end{center}
  \caption{The \sandbox API design shifts general-purpose components out of the
  \sandbox and into the \rpcproxy.}
\label{tab:us-sz-breakdown-sandbox}
\end{table}

Unlike the full \vecdb application, a \sandbox written for \vecdb need only
include support for marshaling a small set of RPCs to fetch its soft-state from
a single service.  More importantly, the design of the \sandbox API places many
general-purpose components in the \rpcproxy, allowing the \sandbox binary to be
small.  \autoref{tab:us-sz-breakdown-sandbox} shows which components the
\sandbox API shifts out of the \sandbox and into the \rpcproxy.

\paragraph{Summary} \Sandboxes can be small because they contain only a small
subset of the functionality of a full microservice, namely initialization, and
because the design of the high-level \sandbox API allows the platform to supply
much of the machinery required for \sandboxes to run and communicate.

\subsection{\Sandboxes are complementary to platforms with fast setup}
\label{ss:eval-blink}

\Sandboxes are a complementary techinque to prior work which reduces setup and
initialization costs. In order to demonstrate the benefit \sandboxes provide to
applications on these platforms we run \imgrecp with Spice~\cite{holmes:spice},
a state-of-the-art serverless snapshot and restore system.

We use Spice to generate a snapshot of an initialized \imgrecp function which
has loaded all necessary Python libraries to perform inference. Then, we invoke
\imgrecp on the same machine, and measure start latency from the point the
function is invoked until it begins performing inference. The input is a 10MB
input image downloaded from a \sigmaos storage service. The \sandbox version of
\imgrecp uses a \sandbox to fetch the input while Spice restores the function.
A good result would show lower start latency for the \sandbox version of
\imgrecp, because the \sandbox can start quickly and begin initialization while
Spice sets up and restores the \imgrecp snapshot.

\begin{figure}
 \includegraphics[width=\linewidth,keepaspectratio]{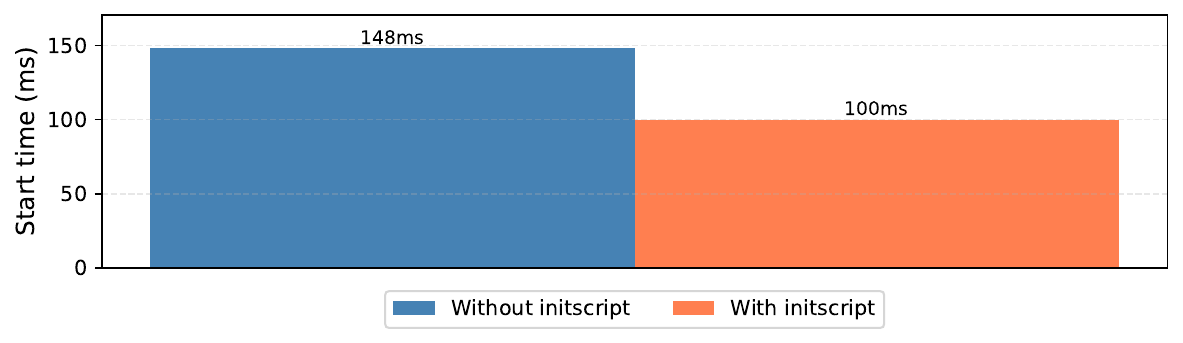}
  \caption{Start latency of \imgrecp restored from a Spice snapshot, without
  and with \sandboxes.}
  \label{fig:blink-start-latency}
\end{figure}

\autoref{fig:blink-start-latency} shows the results. By using \sandboxes,
\imgrecp is able to reduce its start latency by a factor of
\spicespeedup$\times$. This is possible because its \sandbox establishes
connections to the storage service and fetches the input image in parallel with
the Spice restore procedure. The connection setup and download, which take
approximately {\spicedl}ms, overlap with much of the {\spicesetup}ms Spice
setup phase required to get to the first line of function code, and with some
of the {\spiceweightsdl}ms required to load the model weights from the snapshot
into the inference library.

Spice represents a lower-bound on the start latency modern platforms can
provide for serverless applications. Other systems, like Mitosis, incur
additional network latency when starting functions, because they fetch their
contents from a remote snapshot or Zygote. This additional latency could be
used by \sandboxes to perform additional initialization steps and to
further reduce start latency for these systems.

\paragraph{Summary} \Sandboxes are a complementary technique to prior work on
fast setup and initialization. Using \sandboxes together with these systems
further improves cloud application start latency.

\section{Related Work}
\label{s:related}

The main contribution of this work is \sandboxes, a new technique to reduce
end-to-end cloud application cold-start time by overlapping setup and
initialization.  A great deal of prior work has sought to improve cold-start
for applications by reducing setup costs, and some researchers have explored
pushing initialization operations, like fetching input data, into the platform.
\Sandboxes are complementary to this work, and allow platforms to further
reduce application cold-start time. Developers can use \sandboxes' scriptable
interface to specify application initialization to the platform separately from
the rest of the application.  In exchange, the platform can reduce start
latency by overlapping setup and initialization.

This section describes prior techniques and how they relate to
the goal of accelerating cloud application start latency.

\paragraph{Isolation.} MicroVMs~\cite{weiss:firecracker}, lightweight
virtualization techniques like LightVM~\cite{manco:lightvm}, and
serverless-oriented unikernels like Seuss~\cite{seuss:cadden} seek to provide
VM-level isolation for multi-tenant workloads while keeping start times low.

Systems like GVisor~\cite{google:gvisor}, SigmaOS~\cite{sigmaos},
Cntr~\cite{thalheim:cntr}, Particle~\cite{particle}, and RunD~\cite{li:rund}
streamline containers, eliminate OS-level bottlenecks, and offer constrained
container environments to enable fast container creation.

Several systems improve startup performance by sharing a single process to run
mutually distrustful workloads. Faasm~\cite{shillaker:faasm}, Cloudflare
Workers~\cite{cloudflare:workers}, and Sledge~\cite{gadepalli:sledge} rely on
WASM's language-level isolation, whereas Lightweight Contexts~\cite{litton:lwc}
propose OS primitives to enable independent units of execution to share a
process while preserving isolation.

XFaaS~\cite{sahraei:xfaas} starts serverless functions fast with less
isolation.

Faster isolated execution environment creation reduces setup costs and is a
complementary approach to accelerating application cold-start with \sandboxes.
By overlapping initialization and setup, \sandboxes enable developers to speed
up cold-starts even more. Additionally, some lightweight isolation techniques
sacrifice security guarantees in order to reduce setup costs. \Sandboxes make
setup time useful by overlapping it with initialization, which reduces the
pressure on minimizing setup time. This may enable platforms to recover strong
isolation while preserving fast application start times.

\paragraph{Binary download.} FaaSNet~\cite{faasnet:wang} accelerates container
provisioning using a novel peer-to-peer download system to distribute container
images, and AWS Lambda~\cite{brooker:lambda} uses multiple caching layers in
the datacenter to accelerate container loading.

Mitosis~\cite{mitosis:wei} and SigmaOS~\cite{sigmaos} leverage the host OS'
demand-paging to lazily download application binaries.

Poby~\cite{chang:poby} and Mitosis~\cite{mitosis:wei} use specialized network
hardware to accelerate image downloads.

Downloading application binaries is a fundamental cost which adds latency to
application cold-starts. \Sandboxes allow the developer to use this time to
overlap initialization costs before the application starts running.

\paragraph{Runtime initialization.} One approach to reducing setup costs is to
avoid initializing application runtimes on repeated invocations.  Several prior
works, like Catalyzer~\cite{du:catalyzer} and Spice~\cite{holmes:spice}, take
this approach using snapshot and restore techniques to cache pre-initialized
runtimes.  AFaaS~\cite{chai:fork-in-the-road} builds on Catalyzer by using
trees of increasingly function-specific snapshot layers to skip some runtime
initialization steps shared by multiple applications.
GroundHog~\cite{alzayat:groundhog} enables secure reuse of initialized
containers by returning applications to a clean state between invocations using
snapshot and restore.

Some approaches use specialized hardware to share snapshots across machines.
Sabre~\cite{lazarev:sabre} uses hardware-accelerated compression to
efficiently manage snapshot, while Mitosis~\cite{mitosis:wei} leverages RDMA
hardware and a customized kernel to implement a remote-fork primitive.

FaasCache~\cite{fuerst:faascache} avoids cold-starts using
keep-alives to keep warm functions up and running at a scale commensurate to
predicted load.

\Sandboxes are complementary to these techniques which reduce the setup cost of
runtime initialization. As demonstrated in this paper's evaluation
(\autoref{ss:eval-blink}), \sandboxes can speed up applications which rely on
snapshot and restore techniques by using the restore time to initialize the
application.

\paragraph{Multi-container application platforms.}
Kubernetes~\cite{google:kubernetes} supports applications structured as
multiple containers, tied together into a Pod. Applications in a pod
can communicate, are physically co-located and share file system state.

Kubernetes supports InitContainers~\cite{kubernetes:init-containers}, which
start before an application's main container. InitContainers can be used to
store secrets out of reach of the application container, include debugging
tools not built into the application container, delay application start, and
prepare the filesystem for the application.  Kubernetes
Sidecars~\cite{kubernetes:sidecar-containers} are similar to InitContainers,
but do not have to run to completion before the main application container
starts.

\Sandboxes use different isolation technology from the main application
container, which allows \sandboxes to start quickly. The \sandbox API keeps
their compiled binaries orders-of-magnitude smaller than most container images,
which makes them fast to download. Additionally, the API which \sandboxes use
to communicate with the application enables passing of initialization results,
like established connections and downloaded application state, between the
\sandbox and application.

\paragraph{State management.} Faa\$T~\cite{romero:faast} and
Locus~\cite{pu:locus} manage a distributed cache for serverless functions.
BlitzScale~\cite{zhang:blitzscale} uses datacenter GPU training network fabric
to distribute application state when scaling up, and several
systems~\cite{kuchler:dandelion, aws:step, fouladi:gg, li:dataflower,
li:faasflow, yu:follow-the-data, bhasi:kraken} leverage the DAG structure of
serverless workflows to optimize data movement and dynamically provision
resources.

Nu~\cite{ruan:nu}, AIFM~\cite{ruan:aifm}, PLASMA~\cite{sang:plasma} enable
providers to flexibly manage state by asking developers to write applications
with new APIs.

\Sandboxes enable developers to specify how to load
application state when applications start up.

\section{Conclusion}
\label{s:concl}

This paper presented \sandboxes, a novel abstraction which can be used to
accelerate cloud application cold-start.
\sandboxes allow cloud platforms to overlap application setup and
initialization, and transfer initialization results to the application 
efficiently using the \sandbox result transfer API.

\section*{Acknowledgments}

\bibliographystyle{plainnat}
\bibliography{n-str,p,n,n-conf}

%
%
%
\end{document}